# Real-Valued 2-D Direction of Arrival Estimation via Sparse Representation

Zavareh Bozorgasl [1*], Mohammad Javad Dehghani [2]

[1] Department of Electrical and Electronics Engineering, Shiraz University of Technology, Shiraz, Iran
[2] Department of Electrical and Electronics Engineering, Shiraz University of Technology, Shiraz, Iran
[*] z.bozorgasl@sutech.ac.ir

**Abstract:** Despite many advantages of direction-of-arrivals (DOAs) in sparse representation domain, they have high computational complexity. This paper presents a new method for real-valued 2-D DOAs estimation of sources in a uniform circular array configuration. This method uses a transformation based on phase mode excitation in uniform circular arrays which called real beamspace $l_1$SVD (RB-$l_1$SVD). This unitary transformation converts complex manifold matrix to real one, so that the computational complexity is decreased with respect to complex-valued computations–its computation, at least, is one-fourth of the complex-valued case; moreover, some benefits from using this transformation are robustness to array imperfections, a better noise suppression because of exploiting an additional real structure, and etc. Numerical results demonstrate the better performance of the proposed approach over previous techniques such as C-$l_1$SVD, RB-ESPRIT, and RB-MUSIC, especially in low signal-to-noise ratios.

## 1. Introduction

In the last decades, the problem of direction of arrival (DOAs) estimation of electromagnetic waves impinging on antenna arrays has been widely investigated. This important issue in array signal processing is needed to source localisation in many areas such as radio astronomy, surveillance and emergency situations. Due to high-resolution performance of conventional subspace-type DOAs estimation methods such as multiple signal classification (MUSIC) [1] (see the review in [2]), they have dominated in the literature. In the recent years, because of the fact that DOAs of signals are usually very sparse relative to the entire spatial domain, DOAs estimation in sparse representation has attracted great attention [3]-[7]. Using global matched filter (GMF) to explore beamformer snapshots for DOA estimation was carried out by Fuchs [3]. Malioutov *et al.* combined the idea of enforcing sparsity by using $l_1$-norm constraint and singular value decomposition (SVD) of the array output which was in data domain and called $l_1$-SVD [4]. DOAs estimation in the correlation domain instead of the data domain has been used in [5]. Joint 2-D DOA estimation is done in an L-shaped array consisting of a ULA and a sparse linear array [6]. Sparse representation methods have many advantages, e.g., good performance in low SNRs, correlation of signals, and limited number of snapshots, to name but a few.

Although, the mostly used configuration in the literature is uniform linear arrays (ULAs) [8-12], uniform circular arrays (UCAs) have many advantages over that and other antenna array configurations. UCAs provide 360° azimuthal coverage, provide information on elevation angle of arrivals, and also have azimuthally invariant beam pattern because of their symmetric antenna array configuration [13-15]. The mentioned advantages make UCAs a distinctively useful configuration in many applications, such as the context of DOAs estimation. In this paper, azimuth and elevation of received signals, i.e. 2-D DOAs estimation, in a UCA configuration will be achieved by choosing $l_1$-SVD as basis and extend that to 2-D DOAs estimation to find 2-D DOAs. In the body of the paper, the proposed method referred to RB-$l_1$SVD.

Although direction-of-arrivals estimation methods in sparse representation domain have many advantages, they suffer from high computational complexity. In this paper, we propose an efficient way to transform the complex-valued DOAs estimation into a real one which leads to a considerable reduction of the computational burden. This reduction will be done by a factor of at least four. In order to convert complex-valued manifold matrices of UCAs into real ones which is computationally efficient [13], phase mode excitation [14-17] is used, thereby providing matrix-based computations which are real-valued. A beamforming based on phase mode excitation is used to synthesize a beamspace manifold matrix which has azimuthal variation through a centro-Hermitian vector, and has a symmetric amplitude taper [14]. The mentioned taper is dependent on the elevation angle. Moreover, the new approach benefits from beamspace advantages, e.g., it is robust to array imperfections, and also noise suppression is better than its element space equivalent method. Computer simulations are performed to demonstrate the effectiveness of the proposed method.

Throughout this paper, the following notations are used: Lower case bold face letters denote vectors while upper case bold face letters denote matrices. $\mathbb{C}$ denotes the set of complex number. The superscript $(.)^T$ denotes matrix or vector transpose while $(.)^H$ denotes complex conjugate transpose. The notations $||.||_1$, $||.||_2$ and $||.||_F$ stand for the $l_1$ norm, $l_2$ norm and Frobenious norm, respectively. $\hat{x}$ is an estimate of $x$ and $diag(\mathbf{x})$ is a diagonal matrix with $\mathbf{x}$ being its diagonal elements.

The rest of the paper is organized as follows. Section 2 introduces a data model and review of the 2-D $l_1$-SVD (called C-$l_1$SVD [18], where C stands for complex) for UCAs. Section 3 presents a new method for real-valued 2-D DOA estimation based on C-$l_1$SVD. Section 4 provides some numerical simulations demonstrating performance of the proposed method. Finally, conclusions are drawn in Section 5.

## 2. Data model and 2-D $l_1$-SVD DOAs estimation in uniform circular arrays

Consider $K$ narrowband far-field sources emitting plane waves impinging on a uniform circular array of omnidirectional sensors in a 3 dimensional planewave from azimuth angles, i.e., $\phi \in [0, 360)$ which measured down from the z axis, and elevation angles, i.e., $\theta \in [0, 90)$ which measured counterclockwise from the x axis. Fig. 1 shows the UCA geometry in the $xy$ plane, with $N$ equispaced sensors. Arrow shows the received signal direction.

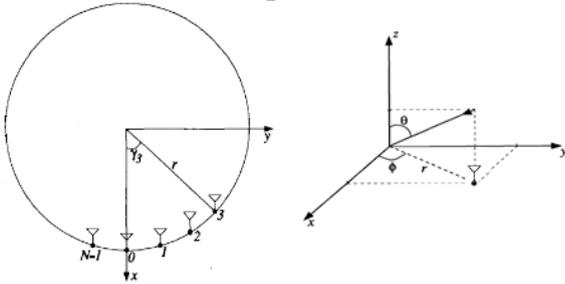

***Fig.1.*** *UCA geometry.*

By employing vector notation for the outputs of the N sensors which placed uniformly on a circle of radius $r$, the array output $\mathbf{x} \in \mathbb{C}^{N \times T}$ is expressed as:

$$\mathbf{x}(t) = \mathbf{A}(\phi,\theta)\mathbf{s}(t) + \mathbf{e}(t), \qquad t = \{t_1, t_2, ..., t_T\}, \qquad (1)$$

where $\mathbf{A}(\phi,\theta) = [\mathbf{a}(\phi_1,\theta_1), \mathbf{a}(\phi_2,\theta_2), ..., \mathbf{a}(\phi_K,\theta_K)]$ is an array manifold matrix, $\mathbf{a}(\phi_i,\theta_i) = \exp(j\xi_i \cos(\phi_i - \gamma_n))$, $n = 0, 1, ..., N-1$ is the steering vector of the source from $(\phi_i, \theta_i)$ which is determined by the geometry of the sensor array, $\gamma_n = \dfrac{2\pi n}{N}$ is displacement of the $n$th element of the array from the x axis, $\xi_i = k_0 r \sin\theta_i$, with the wave number $k_0 = \dfrac{2\pi}{\lambda}$, $\mathbf{s}(t) = [s_1(t), s_2(t), ..., s_K(t)]^T$ is a source vector, and $\mathbf{e}(t) = [e_1(t), e_2(t), ..., e_N(t)]^T$ is an additive white Gaussian noise vector, $t$ indexes the snapshot and T in $t_T$ is the snapshot number.

The observation model in (1) can be written into a more compact form as:

$$\mathbf{X} = \mathbf{A}(\phi,\theta)\mathbf{S} + \mathbf{E}, \qquad (2)$$

where $\mathbf{X} = [\mathbf{x}(t_1), \mathbf{x}(t_2), ..., \mathbf{x}(t_T)]$, $\mathbf{S} = [\mathbf{s}(t_1), \mathbf{s}(t_2), ..., \mathbf{s}(t_T)]$ and $\mathbf{E} = [\mathbf{e}(t_1), \mathbf{e}(t_2), ..., \mathbf{e}(t_T)]$ are the array output, source and noise data matrices, respectively.
After taking SVD of (2) we have:

$$\mathbf{X} = \mathbf{U}_s \mathbf{\Sigma}_s \mathbf{V}_s^H + \mathbf{U}_c \mathbf{\Sigma}_c \mathbf{V}_c^H , \qquad (3)$$

where the $K$ largest singular values, give $\mathbf{U}_s \in \mathbb{C}^{N \times K}$ and $\mathbf{V}_s \in \mathbb{C}^{T \times K}$ as their corresponding singular vectors. Let $\mathbf{X}_{SV} \triangleq \mathbf{X}\mathbf{V}_s$ and also, $\mathbf{S}_{SV} \triangleq \mathbf{S}\mathbf{V}_s$, $\mathbf{E}_{SV} \triangleq \mathbf{E}\mathbf{V}_s$, then a reduced $N \times K$ dimensional signal space will be obtained as follows,

$$\mathbf{X}_{SV} = \mathbf{A}(\phi,\theta)\mathbf{S}_{SV} + \mathbf{E}_{SV} , \qquad (4)$$

Now, suppose we denote the possible received signals DOAs by $(\phi_i, \theta_i)_{i=1}^P$, so the manifold matrix for 2-D DOAs is $\mathbf{A}(\hat{\phi}, \hat{\theta}) = [\mathbf{a}(\phi_1,\theta_1), \mathbf{a}(\phi_2,\theta_2), ..., \mathbf{a}(\phi_P,\theta_P)]$. Thus, we have:

$$\mathbf{X}_{SV} = \mathbf{A}(\hat{\phi}, \hat{\theta})\mathbf{S}_{es} + \mathbf{E}_{SV} , \qquad (5)$$

where the subscript es in $\mathbf{S}_{es}$ stands for element space.
A straightforward extension of $l_1$-SVD [4] is C-$l_1$SVD which yields 2-D DOAs estimation. As a result, the following convex optimization problem which is an $l_1$-norm relaxation equivalent of an $l_0$-norm minimization problem [19-24], gives us $\mathbf{S}_{es}$

$$\min ||\mathbf{s}_{es}^{l_2}||_1 \quad subject\ to\ ||\mathbf{X}_{SV} - \mathbf{A}(\hat{\phi},\hat{\theta})\mathbf{S}_{es}||_2 \le \beta , \qquad (6)$$

where $[\mathbf{s}_{es}^{l_2}]_i = ||\mathbf{S}_{es}(i,:)||_2$ and a choice for $\beta$ is the upper value of $||\mathbf{E}_{SV}||_2$ which has a confidence interval of 99%.

## 3. Real-valued 2-D $l_1$-SVD

We will convert complex computations to real one by using a transformation in a similar manner as [16]. Obviously, circular aperture can excite only a limited number of modes. Let $M$ denotes the highest order mode that can be excited by the aperture at a reasonable strength. There is a rule of thumb to determine $M$ which will be achieved by using Bessel function properties [15]:

$$M \approx k_0 r , \qquad (7)$$

The beamformer for values of $m \geq M$ severely attenuates sources from all directions. For the sake of simplicity, phase mode excitation of the *N* element UCA will be considered. For values of phase mode which $m \leq M$, the array is excited by the normalized beam-forming weight vector:

$$\mathbf{w}_{|m|\leq M}^H = \frac{1}{N}[e^{jm\gamma_0}, e^{jm\gamma_1}, ..., e^{jm\gamma_{N-1}}] = \frac{1}{N}[1, e^{j2\pi m/N}, ..., e^{j2\pi m(N-1)/N}] \quad , \quad (8)$$

which results in the following array pattern

$$f_m^s(\phi,\theta) = \omega_m^H \mathbf{a}(\phi,\theta) = \frac{1}{N}\sum_{n=0}^{N-1} e^{jm\gamma_n} e^{j\xi \cos(\varphi-\gamma_n)} \quad (9)$$

where s in the superscript denotes the sampled aperture.

The array pattern for values of mode order that $|m| \leq N$ is

$$f_m^s(\phi,\theta) = j^m J_m(\xi) + \sum_{q=1}^{\infty}(j^q J_q(\xi)e^{-jg\phi} + j^h J_h(\xi)e^{-jh\phi}), |m| < N \quad (10)$$

where $g$ and $h$ are defined as $g = Nq - m$ and $h = Nq + m$. The first term of (10) is principal term and corresponds to continuous aperture, and the remaining terms arises from continuous aperture sampling.

We can consider sources in the farfield of sensor array (receivers)- which is equivalent of dominating the principal term in the (10) equation-, if the number of array elements satisfies [15]:
$$N \geq 2M \quad , \quad (11)$$

In order to make the transformation from element space to beamspace, we use $\mathbf{F}_r^H$. This transformation will be done by defining a matrix $\mathbf{F}_r^H = \mathbf{W}^H \mathbf{F}_e^H$, where $\mathbf{W}^H$ matrix has centro-Hermitian row. Both of beamformers in this definition, i.e., $\mathbf{F}_r^H$ and $\mathbf{F}_e^H$, synthesize manifolds of size $M' = 2M + 1$. The beamforming matrix $\mathbf{F}_e^H$ is given by:

$$\mathbf{F}_e^H = \mathbf{C}_v \mathbf{V}^H \quad , \quad (12)$$

where
$$\mathbf{C}_v = diag\{j^{-M}, ..., j^{-1}, j^0, j^{-1}, ..., j^{-M}\},$$
$$\mathbf{V} = \sqrt{N}[\mathbf{w}_{-M} \vdots \cdots \vdots \mathbf{w}_0 \vdots \cdots \vdots \mathbf{w}_M] \quad .$$

The beamformer $\mathbf{F}_r^H$ gives a real-valued beamspace manifold $\mathbf{b}(\phi,\theta)$. $\mathbf{F}_r^H$ and $\mathbf{b}(\phi,\theta)$ are defined respectively as (13) and (14):

$$\mathbf{F}_r^H = \mathbf{W}^H \mathbf{F}_e^H = \mathbf{W}^H \mathbf{C}_v \mathbf{V}^H, \quad (13)$$
$$\mathbf{b}(\phi,\theta) = \mathbf{F}_r^H \mathbf{a}(\phi,\theta) \quad , \quad (14)$$

Let $\tilde{\mathbf{I}}$ denotes the exchange matrix, and then any matrix that satisfies $\tilde{\mathbf{I}}\mathbf{W} = \mathbf{W}^*$ can be used in (13). The matrix $\mathbf{W}$ will be chosen as [16]:

$$\mathbf{W} = \frac{1}{M'}[\mathbf{v}(\alpha_{-M}) \vdots \cdots \vdots \mathbf{v}(\alpha_0) \vdots \cdots \vdots \mathbf{v}(\alpha_M)] \quad , \quad (15)$$

where $\mathbf{v}(\phi) = [e^{-jM\phi}, ..., e^{-j\phi}, e^{j0}, e^{j\phi}, ..., e^{jM\phi}]^T$.

Now, by combining the above idea of beamspace with the C-$l_1$SVD method, a new method in beamspace domain will be achieved. By employing beamformer $\mathbf{F}_r^H$ to (5), transformation from element space to beamspace happens and results in:

$$\mathbf{Y} = \mathbf{F}_r^H \mathbf{X}_{SV} = \mathbf{B}(\hat{\phi},\hat{\theta})\mathbf{S}_{bs} + \mathbf{F}_r^H \mathbf{E}_{SV} \quad , \quad (16)$$

where $\mathbf{B}(\hat{\phi},\hat{\theta}) = \mathbf{F}_r^H \mathbf{A}(\hat{\phi},\hat{\theta})$ is a real-valued beamspace matrix with columns of $\mathbf{b}(\hat{\phi}_i,\hat{\theta}_i)$, and bs in $\mathbf{S}_{bs}$ stands for beamspace instead of element space. After partitioning (16), into real and imaginary parts as:

$$\mathrm{Re}(\mathbf{Y}) = \mathbf{B}(\hat{\phi},\hat{\theta})\mathrm{Re}(\mathbf{S}_{bs}) + \mathrm{Re}(\mathbf{F}_r^H \mathbf{E}_{SV}) \quad , \quad (17)$$
$$\mathrm{Im}(\mathbf{Y}) = \mathbf{B}(\hat{\phi},\hat{\theta})\mathrm{Im}(\mathbf{S}_{bs}) + \mathrm{Im}(\mathbf{F}_r^H \mathbf{E}_{SV}) \quad , \quad (18)$$

where $\mathrm{Re}(.)$ and $\mathrm{Im}(.)$ show real and imaginary parts, respectively. By writing (17) and (18) in a compact form, to use all of the data, i.e., not throwing away any useful information, we have:

$$\overline{\mathbf{Y}} = \mathbf{B}(\hat{\phi},\hat{\theta})\overline{\mathbf{S}}_{bs} + \overline{\mathbf{F}_r^H \mathbf{E}_{SV}}, \quad (19)$$

where $\overline{(.)}$ denotes a matrix with both real and imaginary parts, i.e., $[\mathrm{Re}(.) \ \mathrm{Im}(.)]$. The SVD of $\overline{\mathbf{Y}}$ is given by:

$$\overline{\mathbf{Y}} = \underline{\mathbf{U}}_s \underline{\mathbf{\Sigma}}_s \underline{\mathbf{V}}_s^H + \underline{\mathbf{U}}_c \underline{\mathbf{\Sigma}}_c \underline{\mathbf{V}}_c^H \underline{\mathbf{V}}_c, \quad (20)$$

where $\underline{\mathbf{U}}_s \in \mathbb{R}^{N \times K}$, $\underline{\mathbf{\Sigma}}_s \in \mathbb{R}^{K \times K}$, $\underline{\mathbf{V}}_s \in \mathbb{R}^{2K \times K}$, $\underline{\mathbf{U}}_c \in \mathbb{R}^{N \times (N-K)}$, $\underline{\mathbf{\Sigma}}_c \in \mathbb{R}^{(N-K) \times (N-K)}$, and $\underline{\mathbf{V}}_c \in \mathbb{R}^{2K \times (N-K)}$.
It is clear that the computation of (20) requires only real-valued computations.
To obtain a reduced dimensional signal space, let $\underline{\mathbf{Y}}_{SV} \triangleq \overline{\mathbf{Y}}\underline{\mathbf{V}}_s$, then a real-valued nonparametric problem will be obtained:

$$\underline{\mathbf{Y}}_{SV} = \mathbf{B}(\hat{\phi},\hat{\theta})\underline{\mathbf{S}} + \underline{\mathbf{E}}_{SV} \quad , \quad (21)$$

where $\underline{\mathbf{S}} = \overline{\mathbf{S}}_{bs}\underline{\mathbf{V}}_s \in \mathbb{R}^{P \times K}$ and $\underline{\mathbf{E}}_{SV} = \overline{\mathbf{E}}_{SV}\underline{\mathbf{V}}_s \in \mathbb{R}^{N \times K}$.

Finally, solving the following convex optimization problem gives the estimate of $\underline{\mathbf{S}}$:

$$\min || \underline{\mathbf{s}}^{l_2} ||_1 \quad subject\ to\ || \underline{\mathbf{Y}}_{SV} - \mathbf{B}(\hat{\phi}, \hat{\theta})\underline{\mathbf{S}} ||_2 \leq \beta , \quad (22)$$

where $[\underline{\mathbf{s}}^{l_2}]_i = || \underline{\mathbf{S}}(i,:) ||_2$ and a choice for $\beta$ is the upper value of $|| \underline{\mathbf{E}}_{SV} ||_2$ which has a confidence interval of 99%. In fact, all matrices in (22) are real so that solving this problem needs only real computations. We note that an additional SVD has negligible computations because of the small size of (20) and existence of real computations only.

*Lemma*: We claim that $|| \underline{\mathbf{E}}_{SV} ||_2$ has a $\chi^2$ distribution whose degrees of freedom is *NK* upon normalization by the value of variance $\sigma^2 / 2$, for moderate to high SNRs. Noise suppression is better than the element space method.

*Proof*: As we mentioned above, noise is independent and identically distributed complex Gaussian whose common variance is $\sigma^2$. Multiplying Gaussian random matrix by orthogonal matrices will not change the distribution due to the property of orthogonal invariances of Gaussian random matrices. Because of decomposition of data into real and imaginary parts with a common variance of $\sigma^2 / 2$ it is straightforward.

*Remark*. The computational complexity of $\mathbf{C}\text{-}l_1\mathrm{SVD}$ is $O((KP)^3)$, while for RB-$l_1$SVD it is decreased by at least a factor of four. It is higher than the cost of MUSIC, i.e., $O(N^3)$.

## 4. Simulation results

In order to investigate the performance of the proposed method, we have considered some examples and compared results with RB-ESPRIT, RB-MUSIC [16] and C-$l_1$SVD [18] methods.

We show the average root mean-square errors (RMSEs) of the DOA estimates, for different SNR values, in *L* Monte Carlo runs. The RMSEs for azimuth and elevation are computed in (23) and (24), respectively,

$$\mathrm{RMSE}_{azimuth} = [\frac{1}{KL} \sum_{k=1}^{K} \sum_{l=1}^{L} (\hat{\phi}_k^l - \phi_k)^2]^{1/2}, \quad (23)$$

$$\mathrm{RMSE}_{elevation} = [\frac{1}{KL} \sum_{k=1}^{K} \sum_{l=1}^{L} (\hat{\theta}_k^l - \theta_k)^2]^{1/2}, \quad (24)$$

where $(\hat{\phi}_k^l, \hat{\theta}_k^l)$ denotes the estimate of $(\phi_k, \theta_k)$ in the *l*th Monte Carlo run (out of *L* runs) and *K* is number of sources. Consider a UCA configuration with $r = \lambda$, *M* = 6, *N* =13 sensors (number of sensors has to satisfy N > 2*M*), the total number of snapshots *T* = 100, and the uniform grid with 1° sampling. Simulation results of RMSE of azimuth and elevation angles of 3 narrowband uncorrelated sources (without loss of generality) with *L*=200 from ($110.1^0, 35.3^0$), ($120.8^0, 45^0$) and ($170.5^0, 85^0$) are presented in Fig. 2 and Fig. 3, respectively. It can be observed from the Figs. that proposed algorithm (RB-$l_1$ SVD) outperforms the methods of C-$l_1$ SVD, RB-ESPRIT, RB-MUSIC.

To acheive better precision and also reduction of computational complexity, the adaptive grid refinement approach is used which is introduced in [4] (i.e., we applied it to combat the effects of a bias caused by limitation of the estimates into a finte set of grid points), with 0.1° uniform sampling for the refined grid.

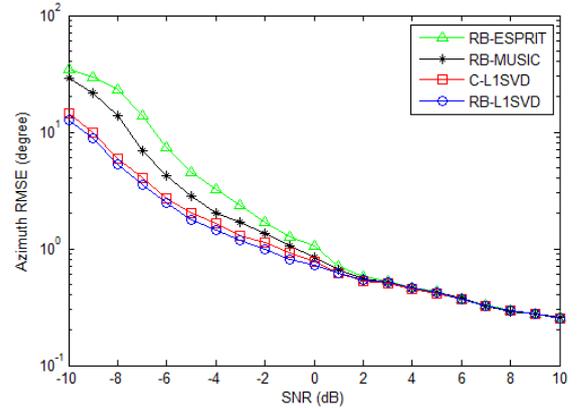

*Fig. 2. RMSE of azimuth estimation of 3 uncorrelated sources from* ($110.1^0, 35.3^0$) , ($120.8^0, 45^0$) *and* ($170.5^0, 85^0$) *angle of arrivals.*

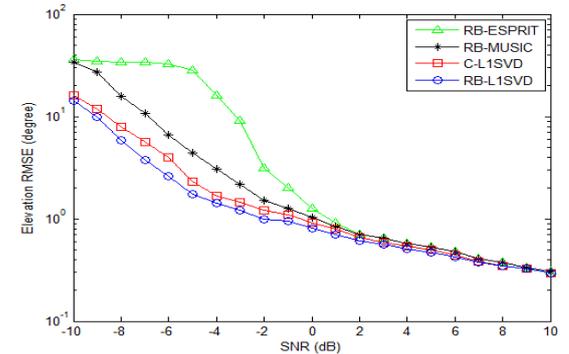

*Fig. 3. RMSE of elevation estimation of 3 uncorrelated sources from* ($110.1^0, 35.3^0$), ($120.8^0, 45^0$) *and* ($170.5^0, 85^0$) *angle of arrivals.*

Fig. 4 shows the resolving capability of closely spaced sources (resolution probability) versus input SNR in 200 Monte Carlo runs for the case of $r = \lambda$, $N = 13$, $T = 100$ and two sources from (200.3, 69.4) and (205.7, 74.5). We claim that the two signals are resolved in a given run, if $\max_{k=1,2} \{|\phi_k - \phi_k|\}$ is smaller than $|\phi_1 - \phi_2|/2$ and $\max_{k=1,2} \{|\theta_k - \theta_k|\}$ is smaller than $|\theta_1 - \theta_2|/2$, where $\phi_k$ and $\theta_k$ stand for the estimated azimuth and elevation DOAs of the *k*th signal, respectively. As it is shown in Fig. 4, the proposed method, i.e., RB-$l_1$SVD, outperforms the subspace-type methods in [16], and also C-$l_1$SVD [18].

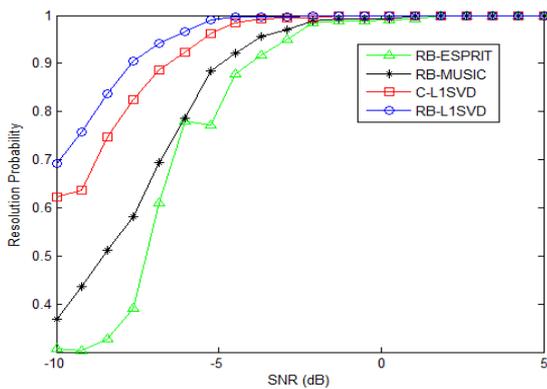

*Fig. 4. Resolution probability versus input SNR for closely spaced sources.*

## 5. Conclusion

We proposed a real-valued 2-D DOAs estimation sparse representation method that uses a beamspace transformation to convert complex-valued computations into real ones, i.e., decreasing computational burden by a factor of at least four. It outperforms other 2-D DOA estimation methods, especially in low signal to noise ratio (SNR) and also it benefits from advantages of beamspace. Simulation results demonstrate better performance of the proposed 2-D DOAs estimation method with respect to RB-MUSIC, RB-ESPRIT and C-$l_1$SVD.

For future studies, we can consider the case of unknown noise statistics, and/or unknown number of sources and/or wideband scenarios. Moreover, extending the problem to other array configurations and considering imperfections such as mutual coupling can be investigated.

## 6. Acknowledgments

Authors would like to thank Dr. Dmitry Malioutov for his great insightful comments.

## 7. References

[1] R. O. Schmidt, "Multiple emitter location and signal parameter estimation," IEEE Trans. Antennas Propag., vol. 34, pp. 276-280, 1986.

[2] H. Krim and M. Viberg, "Two decades of array signal processing research," IEEE Signal Process. Mag., vol. 13, pp. 67-94, Jul. 1996.

[3] J. J. Fuchs, "On the application of the global matched filter to DOA estimation with uniform circular arrays," IEEE Trans. Signal Process.,vol. 49, no. 4, pp. 702-709, Apr. 2001.

[4] D. Malioutov, M. Cetin, and A. S. Willsky, "A sparse signal reconstruction perspective for source localization with sensor arrays," IEEE Trans. Signal Process., vol. 53, no. 8, pp. 3010-3022, Aug. 2005.

[5] P. Stoica, P. Babu, and J. Li, "SPICE: A sparse covariance-based estimation method for array processing," IEEE Trans. Signal Proces.,vol. 59, no. 2, pp. 629-638, Feb. 2011.

[6] J. F. Gu, W. P. Zhu, and M.N.S. Swamy, "Joint 2-D DOA estimation via sparse L-shaped array," IEEE Trans. Signal Process., vol. 63, no. 5, pp. 1171-1182, Jan. 2015.

[7] J. Yin and T. Chen, "A sparse representation-based DOA estimation algorithm with seperable observation model," IEEE Antennas Wireless Propag. Lett., vol. 14, pp. 1586-1589, 2015.

[8] H. C. So, "Source localization: Algorithms and analysis," in Handbook of Position Location: Theory, Practice and Advances, S. A. Zekavat and M. Buehrer, Eds. New York, NY, USA:Wiley-IEEE Press, 2011.

[9] F. Sellone and A. Serra, "A novel online mutual coupling compensation algorithm for uniform and linear arrays," IEEE Trans. Signal Process., vol. 55, pp. 560-573, Feb. 2007.

[10] Md M. Hyder and K. Mahata, "Direction-of-arrival estimation using a mixed $l_{2,0}$ norm approximation," IEEE Trans. Signal Process., vol. 58, no. 9, pp. 4646-4655, Sept. 2010.

[11] B. Liao and S. C. Chan, "Adaptive beamforming for uniform linear arrays with unknown mutual coupling, " IEEE Antennas and Wireless Propag. Lett., vol. 11, pp. 464-467, 2012.

[12] Z. Yang, L. Xie, and C. Zhang, "Off-grid direction of arrival estimation using sparse Bayesian inference," IEEE Trans. Signal Process., vol. 61, no. 1, pp. 38-43, Jan. 2013.

[13] Q. Huang , G. Zhang , and Y. Fang , " Real-valued DOA estimation for spherical arrays using sparse Bayesian learning ," Signal Process., vol. 125, pp. 79-86, 2016.

[14] M. D. Zoltowski, G. M. Kautz, and S. D. Silverstein, "Beamspace Root-MUSIC," IEEE Trans. Signal Process., vol. 41, pp. 344-364, Jan. 1993.

[15] A. Balanis. Antenna theory: analysis and design, Wiley-Interscience, 2005.

[16] C. P. Mathews and M. D. Zoltowski, "Eigenstructure techniques for 2-D angle estimation with uniform circular arrays," IEEE Trans. Signal Process., vol. 42, no. 9, pp. 2395-2407, Sep. 1994.

[17] B. R. Jackson, S. Rajan, B.J. Liao and S. Wang, "Direction of arrival estimation using directive antennas in uniform circular arrays,'' IEEE Trans. Antennas Propag., vol. 63, no. 2, pp. 736-747, Jan. 2015.

[18] Z. Bozorgasl, M. J. Dehghani, "2-D DOA Estimation in Wireless Location System Via Sparse Representation, International Conference on Computer and Knowledge Engineering (ICCKE), Sep. 2014, Mashhad, Iran.

[19] E. Candes, J. Romberg, and T. Tao, "Robust uncertainty principles: Exact signal reconstruction from highly incomplete frequency information," IEEE Trans. Inf. Theory, vol. 52, no. 2, pp. 489-509, Feb. 2006.

[20] D. Donoho, "Compressed sensing," IEEE Trans. Inf. Theory, vol. 52, no. 4, pp. 1289-1306, Apr. 2006.

[21] E. J. Candes and C. Fernandez-Granda, "Towards a mathematical theory of super-resolution," Communications on Pure and Applied Mathematics, vol. 67, no. 6, pp. 906-956, 2014.

[22] G. Tang, B. N. Bhaskar, P. Shah, and B. Recht, "Compressed sensing off the grid, " IEEE Trans. on Information Theory, vol. 59, no. 11, pp. 7465-7490, 2013.

[23] Z. Yang and L. Xie, "Enhancing sparsity and resolution via reweigh ted atomic norm minimization, " IEEE Trans. Signal Process., vol. 64, pp. 995-1006, 2015.

[24] Z. Yang and L. Xie, "Exact joint sparse frequency recovery via optimization methods," IEEE Trans. Signal Process., vol. 64, pp. 5145-5157, 2016.